\documentstyle{mn}

\input{epsf}
\begin{document}

\title[Testing ans\"{a}tze for Quasi-nonlinear Clustering:
the linear APM power spectrum]
{Testing ans\"{a}tze for quasi-nonlinear clustering:
the linear APM power spectrum}

\author[C.M. Baugh and E. Gazta\~{n}aga]
{C.M. Baugh$^{1}$  and  {\rm E. Gazta\~{n}aga}$^{2,3}$
 \\
1. Department of Physics, Science Laboratories, South Road, Durham DH1 3LE
\\
2. CSIC Centre d'Estudis Avancats de Blanes, c/ Cam\'i 
de St. Barbara s/n, 
17300 Blanes, Girona, Spain
\\
3. Department of Physics, Keble Road, Oxford OX1 3RH.}

\maketitle

\def\mpc {h^{-1} {\rm{Mpc}}}
\def\and  {{\it {et al.}}}
\def\rmd {\rm d}
\newcommand{\xibar}{\overline{\xi}}
\newcommand{\beq}{\begin{equation}}
\newcommand{\eeq}{\end{equation}}
\def\spose#1{\hbox to 0pt{#1\hss}}
\def\simlt{\mathrel{\spose{\lower 3pt\hbox{$\mathchar"218$}}
     \raise 2.0pt\hbox{$\mathchar"13C$}}}
\def\simgt{\mathrel{\spose{\lower 3pt\hbox{$\mathchar"218$}}
     \raise 2.0pt\hbox{$\mathchar"13E$}}}

\begin{abstract}
We compare the accuracy of published formulae that transform the 
linear perturbation theory power spectrum into 
the nonlinear regime against the results of 
an ensemble of large N-body simulations, paying attention 
to scales on which the density fluctuations are linear and 
quasi-linear. 
The inverse transformation to obtain the linear 
power spectrum is applied to the APM Galaxy Survey
power spectrum measured by Baugh \& Efstathiou (1993).
The resulting linear spectrum is used to generate the initial density 
fluctuations in an N-body simulation, which is evolved
to match the measured APM amplitude
on large scales. 
We find very good agreement between the final 
power spectrum of the simulation and the measured APM power spectrum.
However, the higher  moments for the particle distribution
only match the ones recovered from the APM Survey on large scales,
$R \simgt 10 \mpc$.
On small scales, $R \simlt 10 \mpc$, the 
APM estimations give smaller amplitudes, indicating
that non-gravitational effects, such as biasing,
are important on those scales. 
Our approach can be used to constrain a model
of how light from galaxies traces the underlying mass
distribution.

\end{abstract}


\section{Introduction}

The growth of density fluctuations can be followed accurately using linear 
perturbation theory only when the density contrast on a given scale is 
much smaller than unity $\delta \rho /\rho \ll 1$ (see for example Peebles 
1980).
As the fluctuations enter the mildly nonlinear regime, $\delta \rho 
/ \rho \sim 1$, analytic approximations or numerical simulations have 
to be used to follow the evolution of the density field, except in cases 
with idealised geometry ({\it e.g.} Bertschinger 1985).

Several comparisons between the results of N-body simulations and the 
predictions of higher order perturbation theory 
have been made recently 
({\it e.g.} Buchert, Melott \& Weiss, 1993, Jain \& Bertschinger 1994).
Baugh \& Efstathiou (1994b) demonstrated that Eulerian 
second order perturbation 
theory gives a good approximation to the evolution of the power 
spectrum in Standard Cold Dark Matter (flat universe, $\Omega=1$, with 
$h=0.5$\footnote{The Hubble constant is given by 
$H_{0} = 100 h {\rm km s}^{-1} {\rm Mpc}^{-1}$}: 
hereafter SCDM) down to scales for which the variance 
$\xibar_2 \sim 1$
correctly predicting a transfer of power from large to small scales. 

An alternative approach for following the evolution of the density field into 
the nonlinear regime has been adopted by Hamilton \and (1991), who 
deduced a transformation between the linear volume averaged correlation 
function and the nonlinear correlation function, with the 
functional form calibrated 
against the results of numerical simulations.
Peacock and Dodds (1994 - hereafter PD) proposed that a similar form 
of the transformation could be applied to the power 
spectrum of density fluctuations.
Jain, Mo and White (1995 - JMW) suggested a correction that improves 
the performance of the Hamilton \and \, and PD formulae when the effective 
slope of the power spectrum, $P(k) \propto k^{n}$ has the value 
$n < -1$.

In Section 2 of this Letter, we compare the predictions of the formulae of 
PD and JMW for the evolution of the shape of the power 
spectrum in an SCDM universe against the results of an 
ensemble of large N-body simulations.

We use these formulae to compute the linear power spectrum corresponding 
to the power spectrum measured for APM Survey galaxies by Baugh 
\& Efstathiou (1993, 1994a) in Section 3.
We use this linear power spectrum to generate the initial conditions 
in an N-body simulation. 
The clustering in the evolved particle distribution is then compared 
with the measurements of the APM power spectrum and the higher order 
moments of counts in cells (Gazta\~{n}aga 1994).
This approach provides a test of our ideas about structure formation,
such as the Gaussianity of the initial conditions and the biasing of 
the galaxy distribution relative to the mass distribution.

\section{Evolution of the CDM power spectrum}

\begin{figure}
\begin{picture}(100, 300)
\put(10,0)
{\epsfxsize=8.truecm \epsfysize=10.5truecm 
\epsfbox[0 0 575 575]{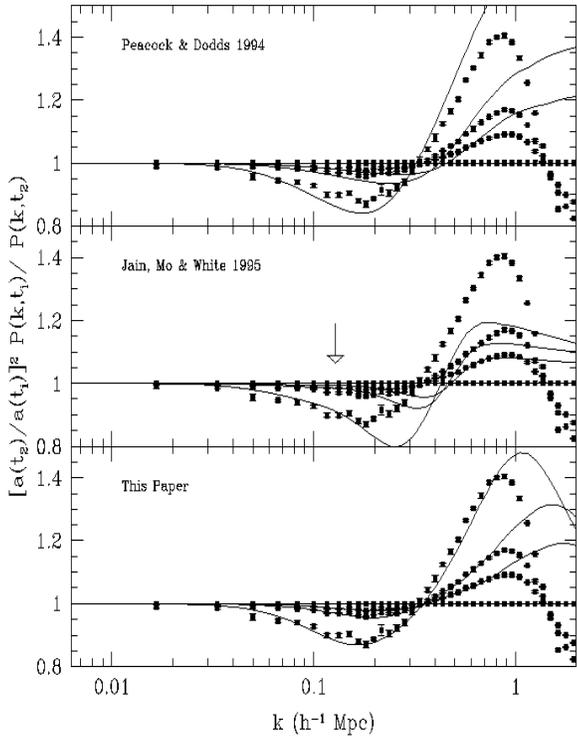}}
\end{picture}
\caption[junk]
{The ratio of the power spectrum at epoch $a_{i}$ to the 
power spectrum at some reference epoch $a_{0}=0.4$ (where $a=1$ corresponds 
to $z=0$), 
with the linear growth scaled out.
The points show the results from an ensemble of large N-body simulations.
The lines show the predictions of the PD and JMW formulae for 
the linear to nonlinear transformation.
The arrow in the middle panel shows the size of the simulation box 
used in the largest volume run of Jain \and.
In the bottom panel, the lines show the fit quoted in the text.}
\label{fig:ratio}
\end{figure}

In this Section we shall examine the CDM power spectrum as an 
example of a scale dependent spectrum.
The nonlinear growth of density fluctuations 
can be studied by comparing the shape of the power 
spectrum of the fluctuations at different epochs.
Following Baugh \& Efstathiou (1994b), we take the 
ratio of the power spectrum at expansion factor 
$a_{i}$ corresponding to time $t_{i}$, to the power 
spectrum at some reference epoch, set by $a_{0}$ and time 
$t_{0}$, scaling out the linear growth factor $\propto a^{2}$:

\begin{equation}
R(k, t_{i}; t_{0}) = \left[ \frac{a_{0}}{a_{i}}\right]^{2} 
\frac{P(k,t_{i})}{P(k,t_{0})} .
\label{eq:ratio}
\end{equation}
This approach takes into account any inaccuracies in the 
initial realisation of the theoretical 
power spectrum (see Baugh \& Efstathiou 1994b). 
We have found that an expansion 
factor of about 3 is necessary to remove transients arising from the Zeldovich 
approximation that is used to set up the initial fluctuations 
(see Baugh, Gazta\~{n}aga \& Efstathiou 1995,
hereafter BGE95), hence our choice 
of $a=0.4$ as  a reference epoch.

The general form of the transformation of linear to nonlinear 
power proposed by PD is given by
\begin{eqnarray}
\Delta^{2} (k_{NL}) &=& f_{NL} [ \Delta^{2}_{L} (k_{L})] \\
k_{L} &=& [ 1 + \Delta^{2}_{NL} (k_{NL}) ]^{-1/3} k_{NL},
\end{eqnarray}
where the subscripts $L$ and $NL$ refer to linear and non linear 
respectively and $\Delta (k) = 4 \pi k^{3} P(k) / (2 \pi)^{3}$ is 
the fractional variance of the density field in bins of $ \ln k$.
The JMW correction requires making a shift in this universal relation:

\begin{equation}
\Delta^{2} (k_{NL})/b(n) =  f_{NL} [ \Delta^{2}_{L} (k_{L})/ b(n)], 
\end{equation}
where the factor $b(n)$ is a function of the spectral index of the 
power spectrum.
JMW obtained the form of this factor $b(n)$ by requiring that their 
transformation reproduced the results of simulations with scale free 
initial conditions.
They then make the assumption that these results can be applied to the 
case of fluctuations that do not have a scale free initial power spectrum, 
such as CDM.
In this instance, an effective spectral index $n_{e}$, 
is defined as the local slope 
of the power spectrum on the scale at which the variance in density 
fluctuations is unity.

Using the fits given by equation (23) of PD 
and equations 5(b) and 7(a) of JMW, 
we can generate a set of evolved power spectra at different epochs.
We set $a=1$ at $\sigma_8=1$ and choose $a=0.40$ to be the reference 
epoch. We interpolate over the power spectra generated at 
$a=0.50, 0.59, 1.00$ to 
form the ratio in equation (\ref{eq:ratio}).
The ratios predicted by the formulae 
are shown as the solid lines in Figure \ref{fig:ratio}.
Also plotted are the same ratios calculated from an ensemble of 
5 N-body simulations, with a box size of $378 \mpc$ and $126^{3}$ 
particles.
The simulations were run with the $P^{3}M$ code described by Efstathiou 
\and (1985).
The errors are the dispersion in the ratios over the five 
simulations in the ensemble. 
We have made no correction to the N-body results to take into account 
aliasing arising from  assignment of particles to the FFT grid, as these 
corrections largely cancel out when the ratio of power spectra is taken, 
due to the relatively  small change in spectral index between different epochs.
The correction for particle discreteness is uncertain (BGE95) and is small 
due to the large number of particles used. 
These effects make some contribution to shape of the N-body curves at 
very high wavenumbers around the particle Nyquist frequency, 
$k \ge 1 h {\rm Mpc}^{-1}$, but do not affect the comparison on large scales.

In Figure \ref{fig:ratio}, we show a fit for the evolution of the 
power spectrum which gives a good 
match to the results of the large box N-body simulations used in this paper.
We have used the same type of fit as JMW, with 
$b(n_{e}) = 1.16 [(3 + n_{e})/3]^{0.5}$ 
where $n_{e}$ is the effective spectral index at each epoch as defined 
by JMW, with $f_{NL}$ given by 

\begin{equation}
f_{NL}(x) = x \left( \frac{1 + a x + b x^{2} + c x^{3} 
+ d x^{3.5} + e x^{4}} 
{ 1 + f x + \left[e/(11.68)^{2}\right] x^{3}}\right)^{1/2} .
\label{eq:myfit}
\end{equation}

We find $a=0.598$, $b=-2.390$, $c=8.360$,
$d=-9.010$, $e=2.895 $ and $f=-0.424$
by matching the power spectrum in the 
simulations at $a=1$, i.e.$\sigma_8=1$.
The accuracy of the fit is better than $5 \%$ over the range 
$0.02 < k < 1.0 h {\rm Mpc}^{-1}$.
Note that as our simulations
do not have the resolution to probe the highly nonlinear 
regime, where we have forced the 
fit to have the asymptotic form $f_{NL}(x) = 11.68 x^{3/2}$ 
when $x \rightarrow \infty $, as used 
by PD. 
Thus our fit does not necessarily perform well on small 
scales, where JMW results are more reliable.

Whilst the agreement of the JMW formula with the simulation results 
on large scales is within the quoted $20 \%$ accuracy, the transformation 
performs less well than either PD or the fit given in this paper.
This is mainly the result of an overprediction of the power at early epochs. 
This could be due to a number of reasons.
JMW used high resolution simulations in order to examine the 
behaviour of the power spectrum in the highly nonlinear regime. 
This was achieved by using a relatively small box size, as indicated 
by the arrow in the middle panel of Figure \ref{fig:ratio}. 
Hence,  Fourier modes of the density field around
 $k \sim 0.2 h {\rm Mpc}^{-1}$ 
do not have modes on larger scales to couple to, with the 
result that the nonlinear evolution on these scales cannot be followed 
accurately.

\section{N-Body realisations of clustering in the APM Survey}

\begin{figure}
\centering
\centerline
{\epsfxsize=8.5truecm \epsfysize=8.5truecm 
\epsfbox{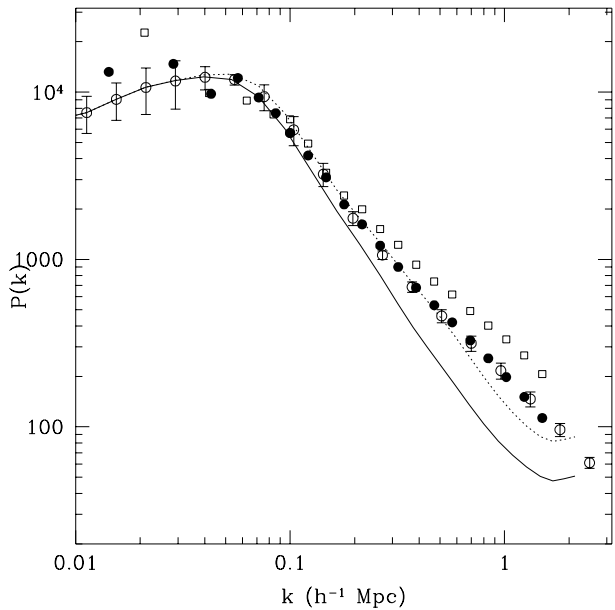}}
\caption[junk]
{Open circles show the APM power spectrum with $1 \sigma$ errors.
The lines show the linear spectrum predicted by the
 JMW (solid) and PD (dotted) formulae. Symbols
show $P(k)$ measured from a N-body
 simulation  with the initial conditions set by the JMW 
(filled circles) and PD (open squares) linear spectra. 
\label{fig:pk}}
\end{figure}

\begin{figure}
\begin{picture}(100,243)
\put(0,3)
{\epsfxsize=8.1truecm \epsfysize=9.0truecm 
\epsfbox[10 150 575 800]{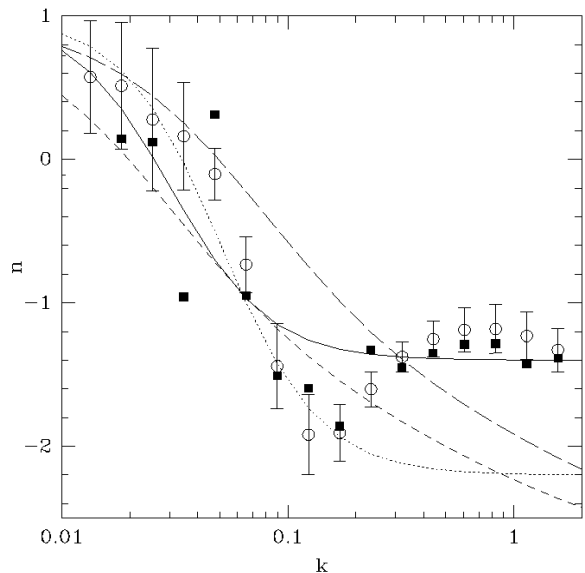}}
\end{picture}
\caption[junk]
{
The logarithmic slope of $P(k)$ as a function of wavenumber estimated 
from the measured APMPK (circles with errors) and from the N-body 
simulation started from initial conditions predicted by the 
JMW formula (filled squares). The errorbars are computed using 
the $1 \sigma$ scatter on the power spectrum. The long (short) 
dashed curve shows the slope in the linear $\Gamma = \Omega h = 
0.5$ ($\Gamma = 0.2$) CDM model.
The solid (dotted) line shows the fit to the measured (linear) APMPK.
}
\label{fig:pkapm2}
\end{figure}

The PD and JMW formulae for the inverse transformation from 
nonlinear to linear power can be used to obtain the linear power 
spectrum that corresponds to the APM Survey power spectrum (APMPK) 
measured by Baugh \& Efstathiou (1993, 1994a).
We make the assumption that there is no bias 
between the galaxy distribution and the mass, {\it i.e.} that the light 
traces the mass.

Figure \ref{fig:pk} shows the APMPK as open circles 
with $1 \sigma $ errors 
obtained by averaging over the APM Survey split up into four zones.
The solid line shows the linear power spectrum predicted by JMW, 
which has a slope of $n \sim -2$,  and 
the dotted line shows that obtained with the PD formula. 
The kink in the linear spectra at $k > 1 h {\rm Mpc}^{-1}$ is an artefact 
due to the form of the transformation formulae.
The simulations that we use are designed to a have particle 
Nyquist frequency at shorter wavenumbers than this and are not 
affected by this feature.
The two linear spectra are quite different. We
find that the simulations evolved from the JMW formula 
give better agreement with the shape of the APMPK, 
confirming that this transformation works best when the power 
spectrum is nearly scale free.
The filled circles show the power spectrum of a $P^{3}M$ simulation with 
$160^{3}$ particles and a box size of $440 \mpc$. 
The simulation has been evolved over $\sim 6$ expansion factors from the 
initial conditions to match the amplitude of the 
variance in spheres of radius $30 \mpc$ given by the APMPK.
Also shown is the result of a smaller simulation, using the PD linear 
power spectrum (open squares), which shows an excess of small scale power 
compared with the APMPK.
PD note that their formula is not expected to work well for such a steep 
spectrum (see also Figure 2 of JMW).
We note that the linear spectrum predicted by our fit is closer to that 
of PD rather than that of JMW; hence a transformation of the type given 
in equation \ref{eq:myfit} which is calibrated against the results of CDM 
simulations does not work well when used with an almost scale free spectrum.

Figure 3 shows $n={\rm d}\ln P(k)/{\rm d}\ln k$, the logarthmic slope,  
estimated from the measured APMPK as a function of wavenumber (open circles).
The dotted line shows the slope of a fit (see below) to the linear APMPK
using the JMW formula. 
The filled squares show the slope of the power spectrum 
after evolution in the N-body simulation, which reproduces 
closely the form of the observed APMPK. 
This Figure clearly shows that the feature in the 
observed APMPK at $k \sim 0.3 {\rm Mpc}^{-1}$ is produced by 
nonlinear evolution, as it is not present in the linear APMPK.
Nonlinear evolution in CDM-like models produces similar behaviour, 
but is insufficient to explain fully the shape of the measured APMPK
(Baugh \& Efstathiou 1994b); the linear APM spectrum is too steep to be fitted by any 
CDM-like model, expressed as a function of the parameter $\Gamma = 
\Omega h$ (Efstathiou \and 1992), which are shown 
as dashed lines in Figure \ref{fig:pkapm2}. 
Efstathiou \and 1990 have shown that it is difficult
to match both the large and small scale shape of the angular correlation  
function measured in the APM  Survey with CDM-like models.
We note however, that the shape of the linear APM power spectrum is similar 
to that predicted in Mixed Dark Matter models (Efstathiou 1995).

To produce the perturbation theory 
predictions in Section 4, we fit the estimated $P(k)$ to a simple
parametric shape:

\beq
P(k)={{C~k^a}\over{\left[1+(k/k_c)^2\right]^b,}}
\label{pkfit}
\eeq 
with a minimum $\chi^2$ fit using the estimated errors. 
Given the additional uncertainties from the APM selection function,
evolution of clustering and value of $\Omega$ (see Gazta\~naga 1995)
we use 2-sigma errors in the APMPK as the error estimation in this fit.
The results are shown in Figure \ref{fig:pkapm2}.
For the non-linear APMPK a fit to the whole range of $k$
gives (solid line): $C \simeq 9.5 \times 10^5$, $k_c \simeq 0.03 \mpc$,
$a \simeq 1$ and $b \simeq 1.2$. 
The fit to the linear JMW  P(k) (dotted line in Figure \ref{fig:pkapm2}) 
is restricted to $k<0.6$ and gives:
$C \simeq 7.0 \times 10^5$, $k_c \simeq 0.05 \mpc$,
$a \simeq 1$ and $b \simeq 1.6$.
The reduced $\chi^{2}$ is much lower for the fit to the linear curve 
than for the nonlinear $P(k)$, showing the difficulty in reproducing the 
interesting features in the APMPK around $k \ge 0.3 {\rm Mpc}^{-1}$ with 
a simple parametric form.
It is interesting to note that although $a$ is a free parameter 
the best fit gives in both cases $a \simeq 1$,  
as predicted by the Harrison-Zeldovich spectrum.

\section{Higher order moments}

We next evaluate the higher order moments in the N-body simulation 
with the same power spectrum as the APM survey.
We use the counts in spherical cells of radius $R$,
to estimate the volume averaged J-order correlation functions
$\xibar_J(R)$, as described in BGE95. 
We concentrate on the higher
order moments in terms of the hierarchical amplitudes 
$S_J \equiv \xibar_J/\xibar_2^{J-1}$.
These quantities can be predicted in perturbation theory  for models 
with Gaussian initial fluctuations which evolve only
under gravity. Bernardeau (1994) has estimated $S_J$ for
the case of a spherical (top-hat window) cell, which 
are given in terms of $J$-order logarithmic derivaties $\gamma_J$
of $\xibar_2(R)$ in the initial conditions. 
These predictions have been tested up
to $J=10$ in N-body simulations (Gazta\~naga \& Baugh 1995, BGE95), 
showing a very
good agreement for scales where $\xibar_2 \simlt 1$.

\begin{figure}
\centering
\centerline
{\epsfxsize=9.truecm 
\epsfbox{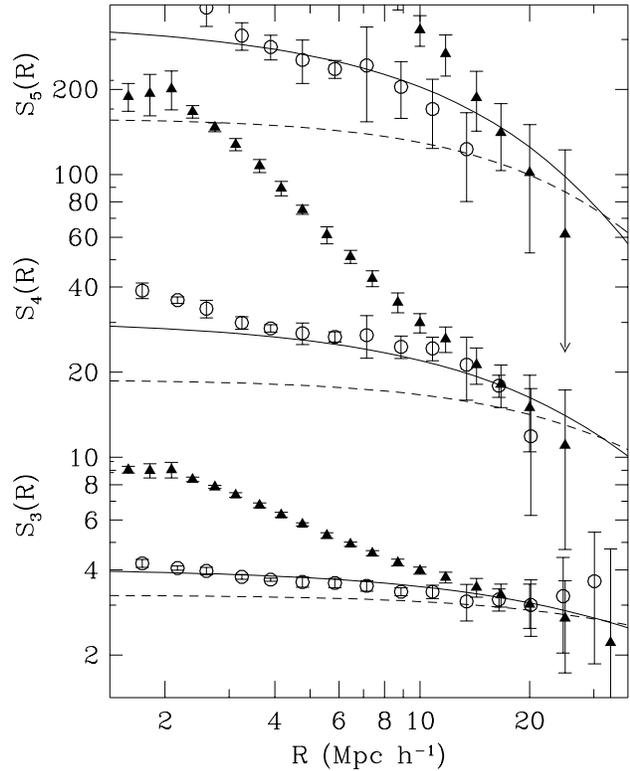}}
\caption[junk]
{The open circles show the APM estimations for $S_3$, $S_4$ and $S_5$.
The closed triangles show the corresponding moments measured in
a large N-body simulation started with the linear APM P(k).
The solid (dashed) line shows the perturbation theory prediction for $S_J$
using linear (non-linear) APM power spectrum.} 

\label{fig:s34}
\end{figure}

In Figure \ref{fig:s34} we compare the perturbation theory predictions for
$S_J$ with $J=3,4,5$ using both the linear (solid line) and non-linear 
(dashed line) shape of $\xibar_2$ estimated 
from the $P(k)$ fits above.
Although we can see in  Figure \ref{fig:s34} 
that these predictions are quite different 
at small scales, they agree well on scales where $\xibar_2 \simlt 1$.
On comparing with the evolved results from the APM like N-body simulations
(closed triangles in  Figure \ref{fig:s34}), we find a good agreement
with perturbation theory predictions 
for scales where $\xibar_2 \simlt 1$, as expected.

\section{Discussion}

The modified transformation from the linear to nonlinear 
regime introduced by JMW seems to work well for the case  of the 
linear APM spectrum presented in Section 3.
However, the transformation is less successful in predicting the 
detailed nonlinear behaviour for the SCDM model. 
The transfer of power in the midly nonlinear regime is of the 
same order of magnitude as the quoted accuracy of the JMW fit.
Whilst we have shown that an improved fit to the behaviour 
of CDM on large scales is possible by constraining the formula to 
match the results of simulations with a large volume, there 
appears to be a fundamental problem in getting a truly universal 
fit that works equally well for scale free and scale dependent spectra, 
with comparable accuracy on large scales to that possible with 
second order perturbation theory (Baugh \& Efstathiou 1994b).

In previous comparisons of perturbation theory predictions for $S_J(R)$ 
with clustering in the APM Survey 
 (e.g. Gazta\~naga 1994, Gazta\~naga \& Frieman 1994,
BGE95) the non-linear shape of $P(k)$ or $\xibar_2$ was used
to calculate the perturbation theory predictions.
This assumes that on large scales the observed shape is not affected by
non-linear biasing or by non-linear gravitational effects. 
Here we drop the latter assumption and use the linear 
$P(k)$, as required in perturbation theory. The comparison of these
improved predictions with the estimations from the APM Survey (open
symbols) is shown in Figure \ref{fig:s34}. 

The APM estimations are the ones presented by Gazta\~{n}aga (1995)
which result from using a simple scaling law to model
the projection effects. 
Although there are some potential
problems with this modelling (Bernardeau 1995), we believe
that these results are accurate (see Gazta\~{n}aga \& Baugh 1996).
The APM amplitudes agree quite well
with the improved predictions on scales $R \simgt 3 \mpc$. 
This is surprising as one would rather expect to find an agreement 
with the fully evolved N-body results, which do not match
these predictions at small scales.
Thus, our analysis indicates the failure of 
at least one of the hypotheses we have used.

We have used $\Omega=1$. This is not very 
important for the 2D to 3D inversion of the APM correlations,
as changing the cosmology only alters the 
overall clustering amplitude slightly and not the shape of 
the correlations
(see Baugh \& Efstathiou 1993, Gazta\~naga 1995).
In the N-body models, a different
value of $\Omega$ would change both the infered initial $P(k)$
and the final $S_J$ in Figure \ref{fig:s34}. We have run some test
models and find that for the APMPK the fitting formulas do not seem
to work that well for $\Omega<1$. 
We find that the spectral index is predicted to be slightly 
more negative on small scales when $\Omega < 1 $, compared with the 
linear power spectrum obtained for $\Omega=1$.
This means slightly larger perturbation theory predictions for $S_J$ and also
more non-linear evolution at high $k$.
We find nevertheless very little difference  
for the final values of $S_J$ for different values of $\Omega$.

In our APM-like simulations we have assumed Gaussian initial
conditions and
in order to infer the linear mass power spectrum from the measured 
galaxy power spectrum we have also assumed that there is no bias between 
the fluctuations in the galaxies and in the underlying density field.
The hierarchical scaling of the higher order moments measured from the 
APM Survey (Gazta\~naga 1994) suggests that there is no relative bias 
between mass and light on large scales.
In addition it is unlikely that 
the scaling could be produced by non-gaussianities or
a particular biasing prescription that 
happens to mimic gravitational growth (Gazta\~naga \& Frieman 1994).
However, the disagreement shown in
Figure \ref{fig:s34} indicates that either of these assumptions
fails on scales $R < 10 \mpc$. 
Some form of non-gravitational effect 
or small scale dependent biasing is necessary. 
Non-gaussian initial
conditions with $S_J < 0$ on small scales, could  account
for the smaller values of $S_J$ in the measurements.
Biasing would also alter the galaxy amplitudes $S_J$ both
 directly and through the change of
the underlying mass power spectrum 
which will lead to a different prediction for the 
linear mass power spectrum. 

{\small 

\section*{Acknowledgements}
We thank George Efstathiou for supplying us with a 
copy of the $P^{3}M$ code and Bhuvnesh Jain for a helpful 
referee's report.
CMB acknowledges receipt of a PPARC research assistantship.
EG acknowledges support from CSIC, DGICYT (Spain), project
PB93-0035 and CIRIT (Generalitat de
Catalunya), grant GR94-8001.

\setlength{\parindent}{0mm}

\bigskip

{\bf REFERENCES} 
\bigskip

\def\refe {\par \hangindent=.7cm \hangafter=1 \noindent}
\def\aj { ApJ, }
\def\aa {A \& A, }
\def\ajs{ ApJS, }
\def\mn { MNRAS, }
\def\apl { ApJ, }

\refe Baugh, C.M., Efstathiou, G., 1993, \mn 265, 145
\refe Baugh, C.M., Efstathiou, G., 1994a, \mn 267, 323
\refe Baugh, C.M., Efstathiou, G., 1994b, \mn 270, 183
\refe Baugh, C.M., Gazta\~{n}aga, E., Efstathiou, G., 1995, \mn 274,
1049 (BGE95)
\refe Bernardeau, F.,  1994 \aa 291, 697
\refe Bernardeau, F.,  1995 \aa 301, 309
\refe Bertschinger, E., 1985 \ajs 58, 39
\refe Buchert, T., Melott, A.L., Weiss, A.G., 1994, \aa 288, 349
\refe Efstathiou, G., Davis, M., Frenk, C.S., White, S.D.M., 1985, \ajs 57, 241
\refe Efstathiou, G., Sutherland, W.J., Maddox, S.J., 1990, Nature, 348, 705
\refe Efstathiou, G., Bond, J.R., White, S.D.M., 1992, \mn 258, 1p
\refe Efstathiou, G., 1995 to appear in proceedings Moriond meeting 
\refe Fry, J.N., Gazta\~naga, E., 1993, \aj 413, 447
\refe Gazta\~{n}aga, E., 1994 \mn 268, 913
\refe Gazta\~{n}aga, E., 1995 \aj 454, 561
\refe Gazta\~{n}aga, E., Baugh, C.M., 1995, \mn 273, L1
\refe Gazta\~{n}aga, E., Baugh, C.M., 1996, in preparation
\refe Gazta\~naga, E., Frieman, J.A., 1994, \apl 437, L13
\refe Jain, B., Bertschinger, E., 1994, \aj 431, 495
\refe Jain, B., Mo, H.J., White, S.D.M., 1995, \mn 276, L25 (JMW)
\refe Hamilton, A.J.S., Kumar, P., Lu, E., Matthews, A., 1991, \aj 374, L1
\refe Peacock, J.A., Dodds, S.J.,  1994, \mn 267, 1020 (PD)
\refe Peebles, P.J.E., 1980 Large Scale Structure of the Universe, Princeton

}
\end{document}